\documentclass[conference]{IEEEtran}
\IEEEoverridecommandlockouts
\usepackage{cite}
\usepackage{amsmath,amssymb,amsfonts}
\usepackage{algorithmic}
\usepackage{graphicx}
\usepackage{textcomp}
\usepackage{xcolor}
\usepackage[utf8]{inputenc}
\usepackage{fourier}
\usepackage{array}
\usepackage{makecell}
\usepackage{subfig}
\usepackage{hyperref}

\def\BibTeX{{\rm B\kern-.05em{\sc i\kern-.025em b}\kern-.08em
    T\kern-.1667em\lower.7ex\hbox{E}\kern-.125emX}}
\begin{document}

\title{QCI Qbsolv Delivers Strong Classical Performance for Quantum-Ready Formulation\\
}

\author{\IEEEauthorblockN{Michael Booth}
\IEEEauthorblockA{\textit{Quantum Computing Inc.} \\
Leesburg, VA, USA \\
}
\and
\IEEEauthorblockN{Jesse Berwald}
\IEEEauthorblockA{\textit{Quantum Computing Inc.} \\
Leesburg, VA, USA \\
ORCID 0000-0003-4741-2427}
\and
\IEEEauthorblockN{Uchenna Chukwu}
\IEEEauthorblockA{\textit{Quantum Computing Inc.} \\
Leesburg, VA, USA \\
ORCID 0000-0002-1311-3827}
\and
\IEEEauthorblockN{John Dawson}
\IEEEauthorblockA{\textit{Quantum Computing Inc.} \\
Leesburg, VA, USA \\
}
\and
\IEEEauthorblockN{Raouf Dridi}
\IEEEauthorblockA{\textit{Quantum Computing Inc.} \\
Leesburg, VA, USA \\
}
\and
\IEEEauthorblockN{DeYung Le}
\IEEEauthorblockA{\textit{Quantum Computing Inc.} \\
Leesburg, VA, USA \\
}
\and
\IEEEauthorblockN{Mark Wainger}
\IEEEauthorblockA{\textit{Quantum Computing Inc.} \\
Leesburg, VA, USA \\
}
\and
\IEEEauthorblockN{Steven P. Reinhardt}
\IEEEauthorblockA{\textit{Quantum Computing Inc.} \\
Leesburg, VA, USA \\
ORCID 0000-0003-4355-6693}
\thanks{Corresponding author: sreinhardt@quantumcomputinginc.com}
}

\maketitle

\begin{abstract} 
Many organizations that vitally depend on computation for their competitive advantage
are keen to exploit the expected performance of quantum computers (QCs) 
as soon as \textit{quantum advantage} is achieved.
The best approach to deliver hardware quantum advantage for high-value problems is not yet clear.
This work advocates establishing \textit{quantum-ready} applications and underlying tools and formulations, 
so that software development can proceed now to ensure being ready for quantum advantage.
This work can be done independently of which hardware approach delivers quantum advantage first.
The quadratic unconstrained binary optimization (QUBO) problem is one such quantum-ready formulation.
We developed the next generation of qbsolv, a tool that is widely used for sampling QUBOs 
on early QCs, focusing on its performance executing purely classically, and deliver it
as a cloud service today.
We find that it delivers highly competitive results 
in all of quality (low energy value), speed (time to solution), and diversity (variety of solutions).
We believe these results give quantum-forward users a reason to switch 
to quantum-ready formulations today, reaping immediate benefits in performance and diversity of solution
from the quantum-ready formulation, preparing themselves for quantum advantage, 
and accelerating the development of the quantum computing ecosystem. 
\end{abstract}


\section{Introduction}

Quantum computers have the potential to enable stunning 
advances, improving the human condition in profound ways -- e.g., saving lives
via faster and more effective drug designs and reducing humanity's impact
on the Earth's environment via better processes for manufacturing fertilizer.
Today, however, no quantum computer has yet delivered \textit{quantum advantage},
i.e., better performance on a real-world problem,
and predictions of when quantum advantage will be achieved range from 1 to 15 years
in the future.
This uncertainty presents a real challenge to quantum-forward organizations wishing 
to exploit the power of QCs as soon as possible.

We are implementing an approach that recognizes the vital role that software will play 
in delivering the performance potential of QCs.
This approach consists of four main threads.  
First, given the uncertainty in when quantum advantage will be delivered and in the details
of potential early QCs (e.g., architecture, number of qubits, and
gates natively implemented) that may deliver quantum advantage, 
application-development formulations and tools must insulate  developers from
that uncertainty, including machine-specific details, to the  extent practical
while still delivering quantum advantage to user applications as soon as QC hardware makes it possible\cite{booth2016abstractions}.
(Obviously the presence or absence of huge QC performance speed-ups cannot be hidden,
but the differences in programming those QCs can be.)
Second, tools should foster the development of hybrid quantum-classical methods that 
use early (small, less-than-robust) quantum processors to deliver the best practical performance.
Third, tools must deliver superior performance now; obviously this will not be the thousand-fold 
or million-fold speed-ups that people expect from QCs, but convincing quantum-forward
organizations to adopt quantum-ready applications and tools requires immediate benefit be delivered.
Lastly, achieving these goals in a fully general way appears daunting in the extreme, so 
narrowing the focus to a particular type of quantum-ready problem type is probably necessary
to achieve near-term success.

One quantum-ready formulation is the quadratic unconstrained binary optimization (QUBO) problem,
also known as the transverse-field Ising model to physicists, 
the binary quadratic problem to operations researchers,
and the probabilistic graphical model to statisticians.
The QUBO has much to recommend it.  
It solves a wide variety of discrete combinatorial optimization problems effectively \cite{kochenberger2014unconstrained},
maps  directly to annealing-based QCs \cite{dash2013note} and effectively to gate-model QCs,
and has a  rich and expanding reservoir of research behind it\cite{lucas2014ising, goddard2017will}.
QUBO formulations have been used for QC-targeted applications 
for graph partitioning for molecular dynamics simulations\cite{mniszewski2016graph},
for protein design\cite{mulligan2019designing}, 
for reducing traffic flow\cite{neukart2017traffic},
and for optimizing workflows within factories\cite{irie2019quantum}.
While it has many strengths, the QUBO requires careful use, as the tuning of Lagrange multipliers, balancing numerical values so that feasible solutions have better energies than non-feasible solutions, complicates getting good results.
Methods have been proven for executing QUBOs, on annealing-based QCs\cite{rosenberg2016building, booth2017qbsolv}, 
even QUBOs bigger than the quantum processor, and on gate-model QCs 
via the Quantum Approximate Optimization Algorithm (QAOA)\cite{farhi2014quantum}, so the quantum-readiness
of the QUBO formulation is established.
In sum, there is considerable evidence the QUBO formulation can play an effective role in solving
discrete combinatorial optimization problems in a quantum-ready way.
\begin{table}[htbp]
\caption{Selected MQlib Instances}
\begin{center}
\begin{tabular}{|c|r|r|c|}
\hline
\thead{Instance \\ name} & \thead{Number of \\ variables} & \thead{Number of \\ nonzeros} & \thead{Density} \\
\hline
g000283 & 3,364	& 406,237	& 0.0718 \\
g000377 & 3,398 &	3,966 &	0.0007 \\
g000421 & 2,034	& 7,756	& 0.0038 \\
g000432 & 2,153	& 250,371 &	0.1081 \\
g000476 & 8,000	& 64,177 &	0.0020 \\
g000495 & 5,438	& 3,151,321	& 0.2132 \\
g000503 & 5,046	& 5,695,279	& 0.4474 \\
g000524 & 2,218	& 336,856 &	0.1370 \\
g000644 & 10,000 &	79,778 &	0.0016 \\
g000788 & 2,342	& 2,404,777	& 0.8772 \\
\hline
g000802 & 3,956	& 1,034,494	& 0.1322 \\
g000969 & 2,453	& 2,581,494	& 0.8584 \\
g000989 & 2,319	& 2,318	& 0.0009 \\
g001086 & 3,706	& 10,898	& 0.0016 \\
g001269 & 2,294	& 4,588	& 0.0017 \\
g001327 & 2,318	& 796,617 &	0.2966 \\
g001337 & 2,850 &	208,125	& 0.0513 \\
g001345 & 5,066	& 9,530,811	& 0.7429 \\
g001393 & 3,938	& 6,412,304	& 0.8272 \\
g001469 & 2,412	& 1,345,896	& 0.4629 \\
\hline
g001581 & 2,383	& 2,447,660	& 0.8624 \\
g001651 & 5,819	& 16,452,342 &	0.9719 \\
g001883 & 6,831	& 13,662 &	0.0006 \\
g001913 & 3,865	& 5,635,669	& 0.7547 \\
g002034 & 2,528	& 1,119,904	& 0.3506 \\
g002204 & 5,368	& 6,318,136	& 0.4386 \\
g002207 & 2,677	& 2,665,638	& 0.7442 \\
g002300 & 5,038	& 11,889,413 &	0.9370 \\
g002312 & 6,395	& 3,830,605	& 0.1874 \\
g002332 & 3,181	& 1,117,657	& 0.2210 \\
\hline
g002370 & 3,884	& 6,327,201	& 0.8391 \\
g002440 & 2,242	& 110,979	& 0.0442 \\
g002512 & 4,731	 & 1,386,782 &	0.1239 \\
g002527 & 5,378	& 8,507,996	& 0.5884 \\
g002563 & 6,279	& 3,705,823	& 0.1880 \\
g002569 & 2,815	& 4,162	& 0.0011 \\
g002586 & 2,079	& 345,114	& 0.1598 \\
g002600 & 2,432	& 2,503,793	& 0.8470 \\
g002898 & 2,041	& 1,795,003	& 0.8622 \\
g003059 & 3,447	& 813,960	& 0.1370 \\
\hline
g003198 & 3,972	& 5,850,724	& 0.7419 \\
g003215 & 2,206	& 2,272	& 0.0009 \\
imgseg\_216041 & 7,724 &	11,689	& 0.0004 \\
imgseg\_376020 & 7,455	& 13,682	& 0.0005 \\
p7000\_2 & 7,001	& 19,505,601 &	0.7960 \\
\hline
\end{tabular}
\label{tbl:MQlib_instances}
\end{center}
\end{table}

Our approach respects another important constraint -- commercial viability.  
We are implementing QCI's Mukai middleware with a broad user base and a software
architecture in mind, knowing that today's version will morph 
in light of tomorrow's better understanding of both quantum hardware and software.
We have a persistent focus on performance, notably the sampling core of QCI qbsolv 
but also efficient interfaces in terms of execution time and memory consumption.
And, we deliver a discrete constrained-optimization sampler in the cloud, running purely classically today
and quantum-ready for the future, and delivering superior performance today. 
The approach is also cost effective; the test runs in this paper were run on virtual machines
in the cloud with a list price less than five dollars per hour.
This makes it practical
for quantum-forward users to shift development and execution of applications requiring the 
speed of future QCs to a quantum-ready software base today.

\section{Implementation and Experimental Design}

\subsection{MQlib and Selected Instances}
MQlib \cite{dunning2018works} is a collection of 3298 QUBOs 
(or maximum cut problems readily converted to QUBOs) 
and 37 QUBO-sampling heuristics open-sourced in 2015.
In addition to that aspect of their work, Dunning et al. studied the execution behavior of each
instance on the heuristics to derive suggested runtimes that distinguished well-performing heuristics
from poorly performing ones.
D-Wave personnel \cite{dwave2020hss} selected 45 \textit{hard} MQlib instances, defined as 
having the maximum suggested runtime of 1200 seconds and only one heuristic achieving the
best reported solution, leaving aside any instances with more than 10,000 variables.
We used the same set of 45 instances, listed in Table \ref{tbl:MQlib_instances}, 
to make our results more easily comparable and to build
consensus for a community-supported benchmark.

\subsection{QCI Qbsolv Implementation}
The MQlib QUBOs were solved with QCI qbsolv, a component of the quantum-ready QCI Mukai software stack.
Derived from the open-source qbsolv\cite{booth2017qbsolv}, QCI qbsolv has been reimplemented 
to deliver exceptional performance (quality, speed, and diversity of results) 
from highly parallel classical processors, exploiting advanced tabu search techniques.  
For these runs, we requested QCI qbsolv to favor quality of results (lowest energy) over fastest
runtime, and return the best 700 answers obtained. 
We ran each problem 5 times and use the results of the run gaining the best energy value
for evaluation.

\subsection{Computer Specs}
For these runs, we executed QCI qbsolv in its current typical setting, an AWS instance visible 
via a Python API exercising a REST API.  
The AWS instance was a c5.24xlarge Linux instance, including 96 Intel Xeon cores running 
at 3.0 GHz (up to 3.4GHz via Turbo Boost) and 192 GiB of RAM.  
QCI qbsolv was given a time limit of 1200 seconds for all problems, equaling the time limit
used in the MQlib runs.  
These results were obtained on Mukai 1.1.

\begin{figure}
    \centering
    \subfloat[Relative delta energy, sorted by increasing relative delta energy]{\includegraphics[width=\columnwidth]{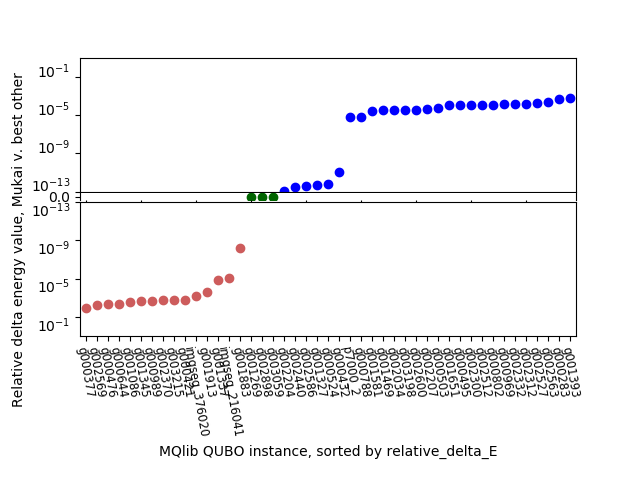}}
    \hspace{0mm}
    \subfloat[Relative delta energy, sorted by increasing problem size]{\includegraphics[width=\columnwidth]{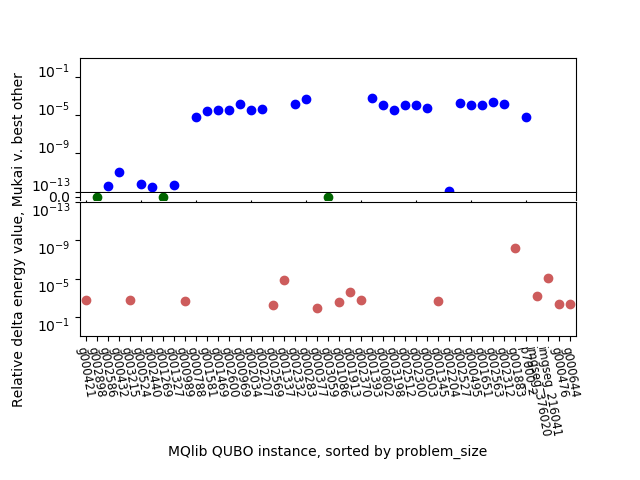}}
    \hspace{0mm}
    \subfloat[Relative delta energy, sorted by increasing problem density]{\includegraphics[width=\columnwidth]{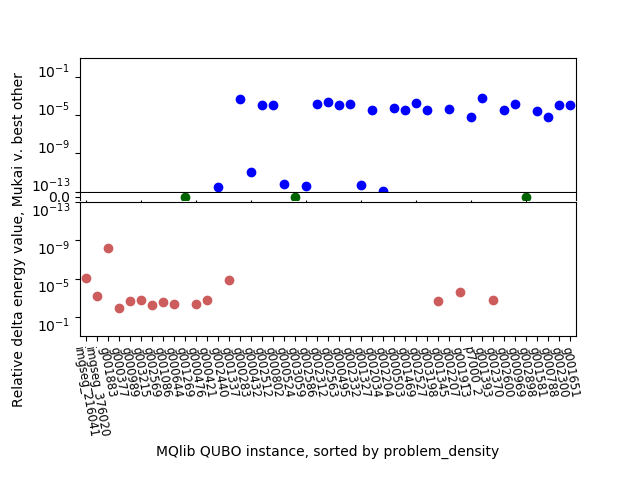}}
    \caption{Relative delta energy, with instances sorted by 3 criteria}
    \label{fig:relative_delta_E}
\end{figure}

\subsection{Performance Metrics}
Measuring the performance of a quantum computer is quite different from measuring
the performance of a classical computer, in that the QC exploits probabilistic 
quantum effects and hence the QC's results are, in general, nondeterministic\cite{mcgeoch2019principles}.
The QUBO form represents NP-hard optimization problems, real-world instances 
of which will be prohibitively expensive to solve exactly, and hence 
the quality of the solution will typically vary from run to run due 
to the probabilistic nature of the QC.
Quality of solution, often stated for the QUBO as best or lowest energy,
is one dimension of QC performance that will be important, 
and it is the dimension that is the first focus of this work.
The second dimension we focus on is the execution time required to obtain those 
low-energy solutions.
Third is the diversity of the best or \textit{elite} solutions, 
often measured pair-wise by the Hamming distance.
The relative importance of these 3 dimensions is application dependent.  
In a case where the selected solution will be realized a great many times, 
such as financial optimization,
even a slightly better solution may be high value.
In another case, 
such as changing a logistics schedule to respond to a dynamic situation,
fixed time to respond to an external stimulus may mean that execution time
below some limit may be critical.
In yet another case, for example protein design,
delivering many possible solutions to be compared against other
criteria may be most important.

\section{Results and Discussion}

\subsection{Best Energy}
Fig. \ref{fig:relative_delta_E} concisely illustrates the energy results we obtained.  
For Fig. \ref{fig:relative_delta_E}(a), for each MQlib instance, we determined whether the best MQlib heuristic or QCI qbsolv
got the best energy value and then calculated the \textit{relative delta energy}, defined as the difference
between the two energy values divided by the best MQlib heuristic value.  
If QCI qbsolv obtained the best energy value, the relative delta energy appears in the upper half
of the plot in \textcolor{blue}{blue}; if the best MQlib heuristic won, the relative delta energy value
appears in the bottom half of the plot in \textcolor{red}{red}. 
If the best MQlib heuristic and QCI qbsolv tied, the point appears on the x-axis in \textcolor{green}{green}. 

We plot the relative energy differences on a log scale so that all the differences are visible.
Some readers may question whether to display relative differences as small as 10\textsuperscript{-13}, but 
the real-world value of a QUBO-solution energy difference is hard to judge;  
e.g., improving the energy of a folded protein by a slight amount 
might enable it to work better in unforeseen ways. 
We document the energy differences and leave assignment of real-world value to subject-matter experts solving
real-world problems.

In Fig. \ref{fig:relative_delta_E}(b), we sort the instances by increasing problem size, to see whether any patterns exist
for which problems QCI qbsolv samples most effectively.  
Problem size, within the range of these problems, appears to have little effect on QCI qbsolv's energy performance relative to
the MQlib heuristics.
We note that we are targeting QCI qbsolv to effectively sample problems well above 50,000 variables,
so the problem-size range exercised here is at the low end of our target.

\begin{figure}
    \centering
    \subfloat[Sampling time, sorted by increasing relative delta energy]{\includegraphics[width=\columnwidth]{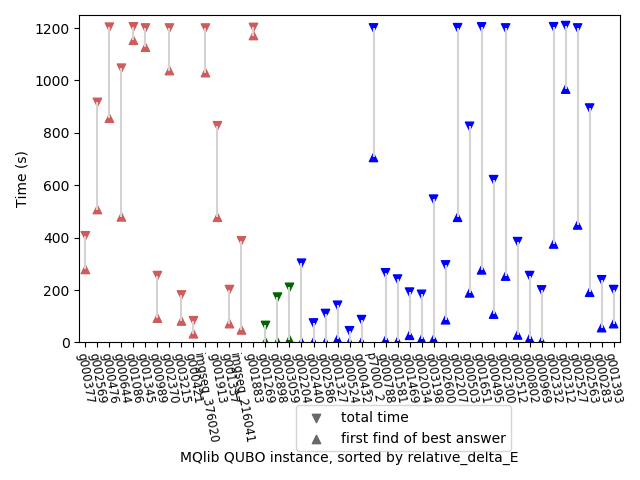}}
    \hspace{0mm}
    \subfloat[Sampling time, sorted by increasing problem size]{\includegraphics[width=\columnwidth]{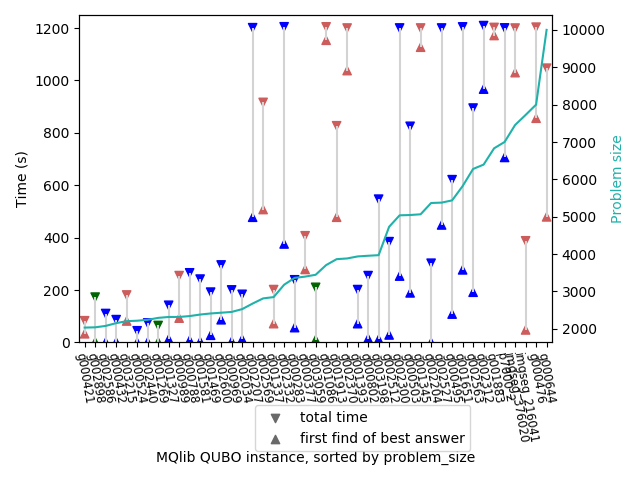}}
    \hspace{0mm}
    \subfloat[Sampling time, sorted by increasing problem density]{\includegraphics[width=\columnwidth]{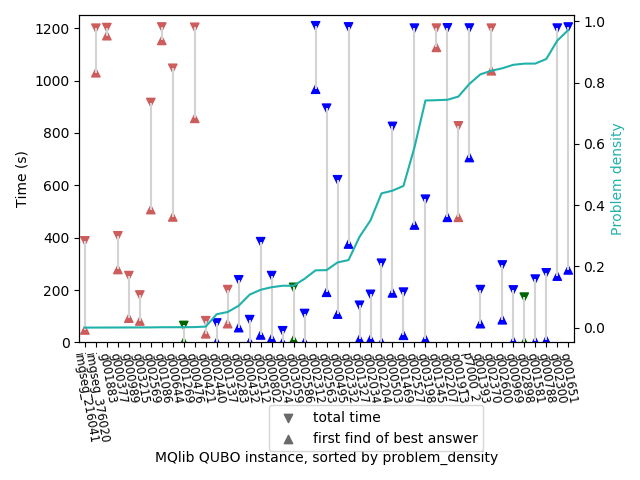}}
    \caption{Sampling time, with instances sorted by 3 criteria}
    \label{fig:time}
\end{figure}

Fig. \ref{fig:relative_delta_E}(c) compares the MQlib heuristics and QCI qbsolv's relative energy differences
to problem density, and here a modest pattern does emerge, where the best MQlib heuristic is often better for
sparser problems but QCI qbsolv is usually superior for denser problems.
We have focused closely on the performance of solving sparse problems, so this is somewhat surprising and
merits further investigation.

\subsection{Time to Solution}
Fig. \ref{fig:time} illustrates, for the runs that produced the best energy values documented in Fig. \ref{fig:relative_delta_E}, the time taken by QCI qbsolv.  
For each instance, two times are plotted, the first ($\bigtriangleup$, at the bottom of the line) 
denoting the time when QCI qbsolv
first found the best answer, and the second ($\bigtriangledown$, at the top of the line) 
denoting the time when QCI qbsolv stopped
searching for a better answer and ended.
The instances are colored the same as Fig. \ref{fig:relative_delta_E}(a) so that patterns of energy vis-a-vis
time may be visually apparent.
In Fig. \ref{fig:time}(b,c) the problem size or problem density, respectively, is
plotted for visual reference.

Fig. \ref{fig:time}(a) sorts the instances by the same order as Fig. \ref{fig:relative_delta_E}(a),
i.e., increasing relative delta energy.
Somewhat surprisingly, there appears to be little correlation between how the energy quality of the
solutions QCI qbsolv finds and how long it requires to find those results.
We see plenty of both red- and blue-capped lines that are relatively short. 
Not surprisingly, for the red-capped lines representing the problems on which QCI qbsolv did relatively
poorly, it took more time, which matches what we want in a QUBO solver -- in general, as long as there is a good
prospect of better answers, keep trying.
We interpret the several blue-capped lines in the right-most x-axis positions that hit the 1200s timeout
as problems where QCI qbsolv is doing relatively well and yet believes it can do still better, 
so keeps trying.

\begin{figure}
    \centering
    \includegraphics[width=65mm]{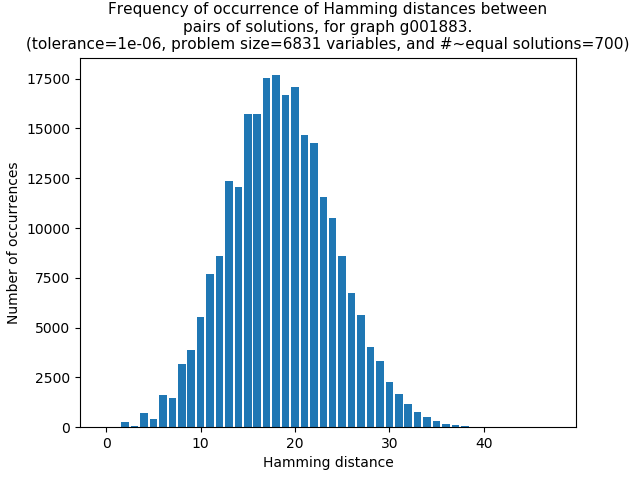}
    \caption{Normal distribution of solution distances.}
    \label{fig:g001883-normal-hist}
\end{figure}

Turning to sampling time sorted by problem size, as shown in Fig. \ref{fig:time}(b), we again see little
pattern regarding on what instances QCI qbsolv performs best.
Red- and blue-capped lines appear roughly randomly distributed through problem size.

\begin{figure}
    \centering
    \subfloat[Number of distinct elite solutions, sorted by increasing relative delta energy]{\includegraphics[width=\columnwidth]{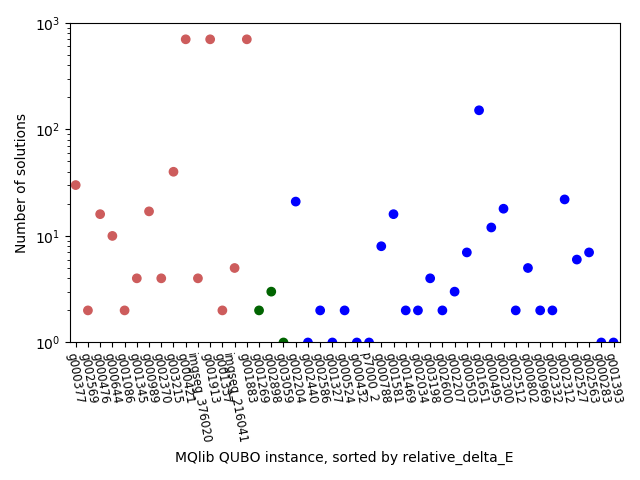}}
    \hspace{0mm}
    \subfloat[Number of distinct elite solutions, sorted by increasing problem size]{\includegraphics[width=\columnwidth]{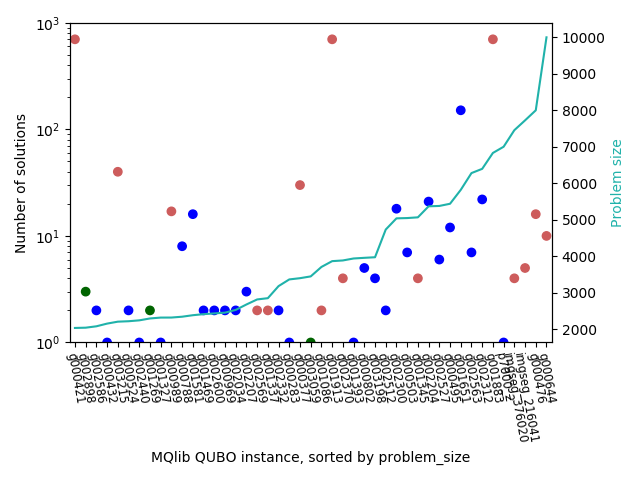}}
    \hspace{0mm}
    \subfloat[Number of distinct elite solutions, sorted by increasing problem density]{\includegraphics[width=\columnwidth]{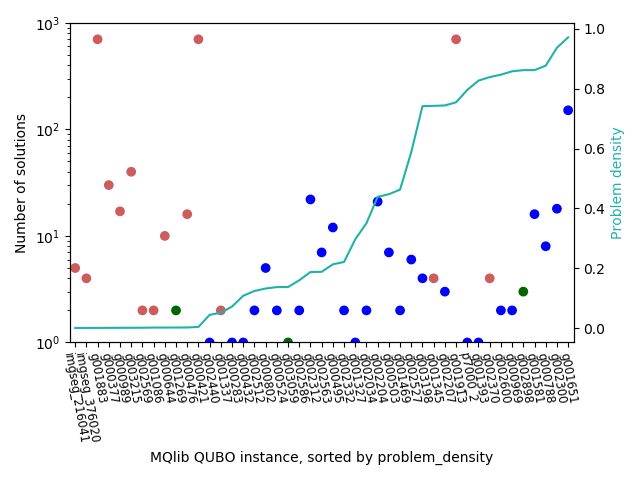}}
    \caption{Diversity (number of distinct elite solutions), with instances sorted by 3 criteria}
    \label{fig:diversity_1}
\end{figure}

In Fig. \ref{fig:time}(c) we see sampling time sorted by problem density and detect a modest pattern
again related to sparse problems, which when coupled with Fig. \ref{fig:relative_delta_E}(c),
indicates that QCI qbsolv is not sampling sparse problems particularly well 
and is taking more time than ideal to do so.

We deduce from the numerous long vertical lines, denoting QCI qbsolv executing for quite some time after
having first found the best answer it found in that run, that there is room for improvement in 
QCI qbsolv's heuristic for deciding that it is not making progress.
(Note again that we configured QCI qbsolv for these runs to find the best energy at some sacrifice of
execution time, so we expect some degree of time overrun.)

\subsection{Diversity}
As noted above, diversity of \textit{elite} solutions, 
i.e., those with energy equal (or nearly so) to the best solutions, 
is a valuable attribute of both a constrained-optimization sampler and a QC.
We are not aware of accepted common metrics for representing diversity beyond the obvious use
of Hamming distance to denote the difference between a pair of solutions.
As noted above, all of our diversity calculations use a tolerance in the relative delta energy
between a solution and the best solution found; for these calculations, that value was
set to $10^{-6}$.

A simple metric is the number of elite solutions, which we plot in  
Fig. \ref{fig:diversity_1} per instance, sorted respectively
by relative delta energy, problem size, and problem density. 
We observe that the number of elite solutions shows no discernible correlation by any of those criteria,
though unfortunately the most elite solutions (the maximum of 700) were found for 3 instances for which
QCI qbsolv did not find the best known energy.

\begin{figure}
    \centering
    \subfloat[Hamming distance information for g001651 solutions]{\includegraphics[width=\columnwidth]{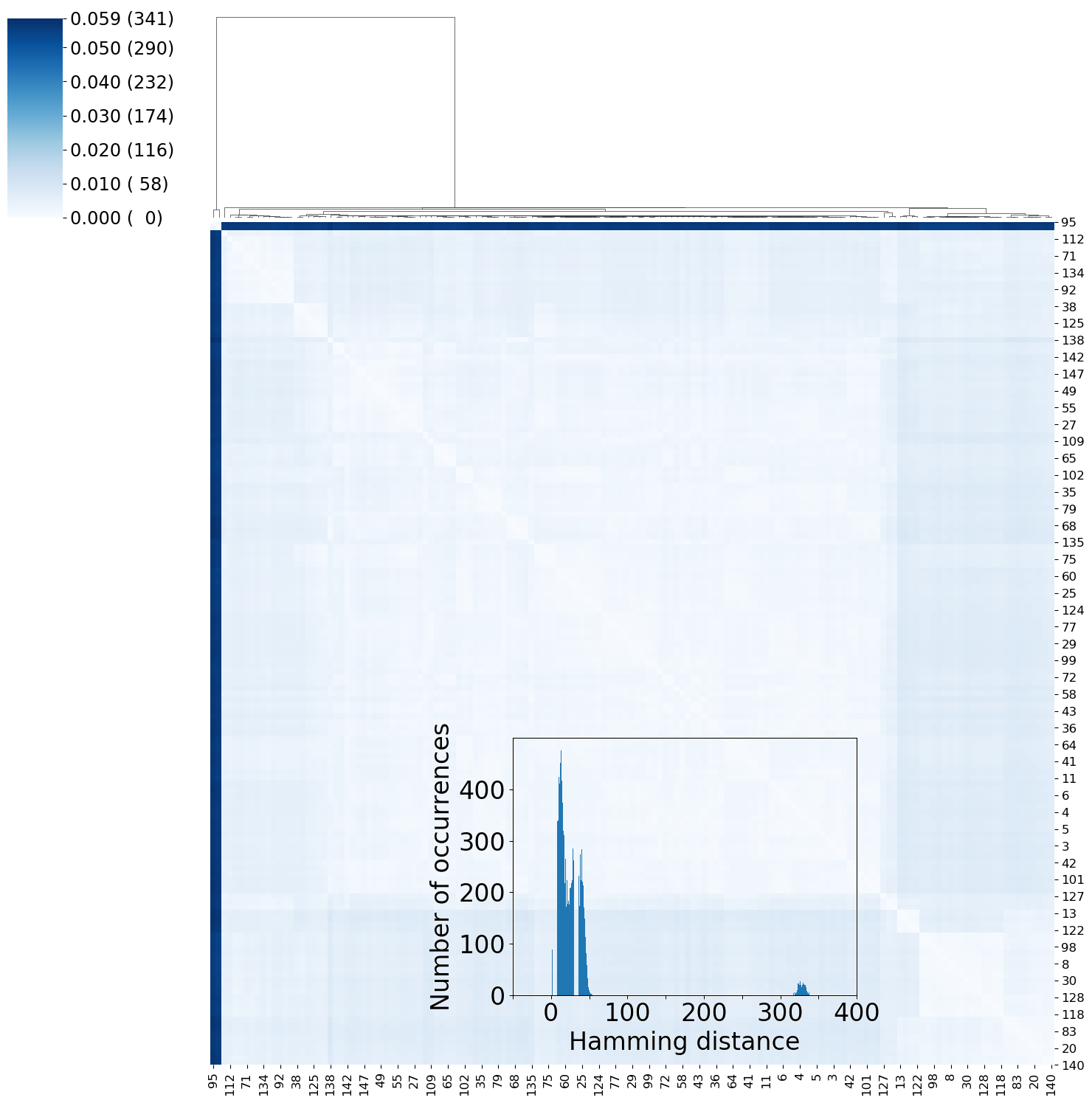}}
    \hspace{0mm}
    \subfloat[Hamming distance information for g002312 solutions]{\includegraphics[width=\columnwidth]{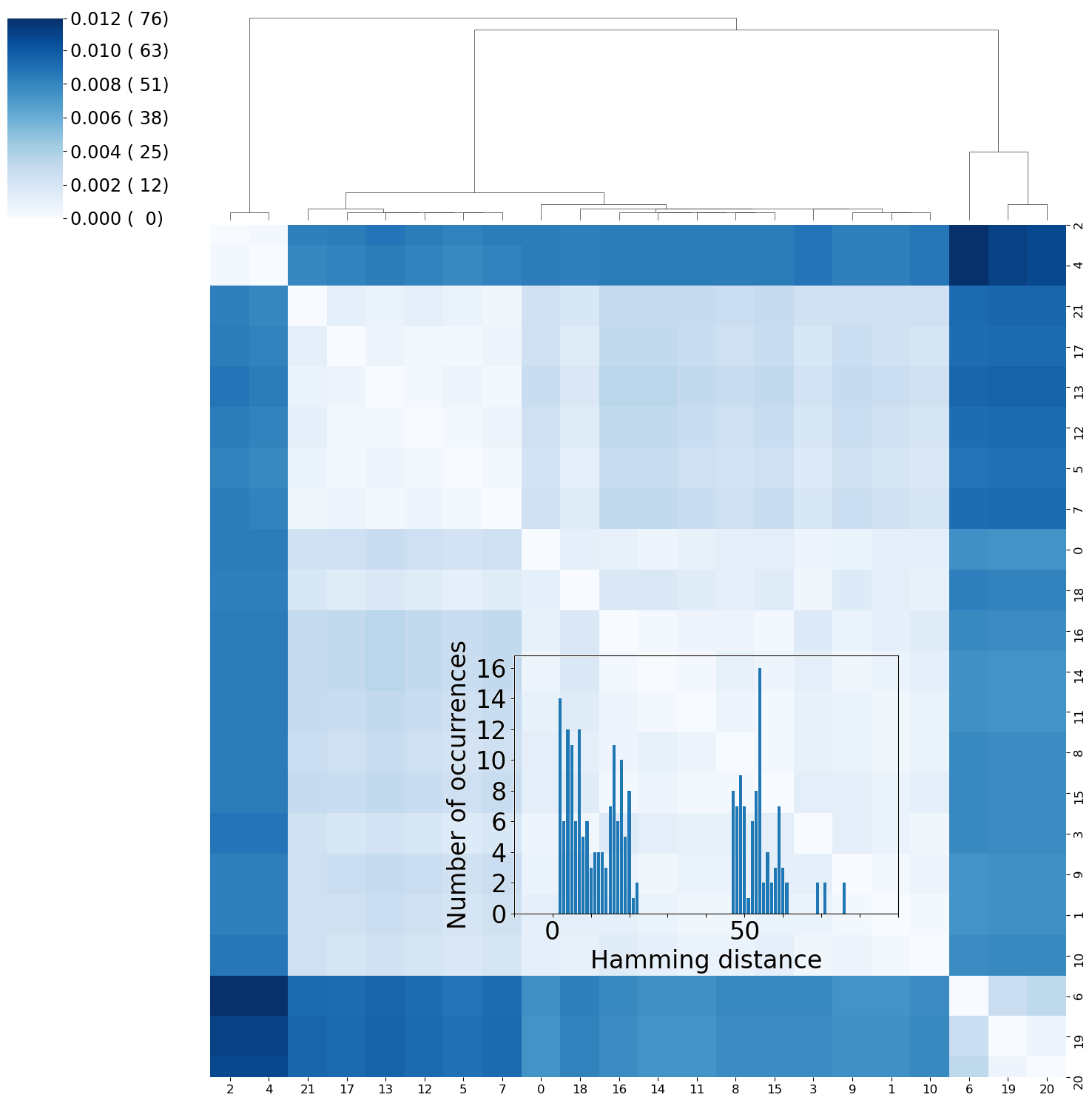}}
    \caption{Hamming distance information for solutions of two problem instances. 
    (a) Most solutions are found very close together, with another cluster a Hamming distance of over 350 away. Within the large cluster the distribution is uniform and is aggregated into a single cluster early in the hierarchical clustering algorithm. This large cluster merges late with the small cluster. (b) The trimodal nature of the solutions is clear in the distance matrix and dendrogram, where the regions merge late in the dendrogram. }
    \label{fig:diversity_2}
\end{figure}

\begin{figure}
    \centering
    \subfloat[Hamming distance information for g000495 solutions]{\includegraphics[width=\columnwidth]{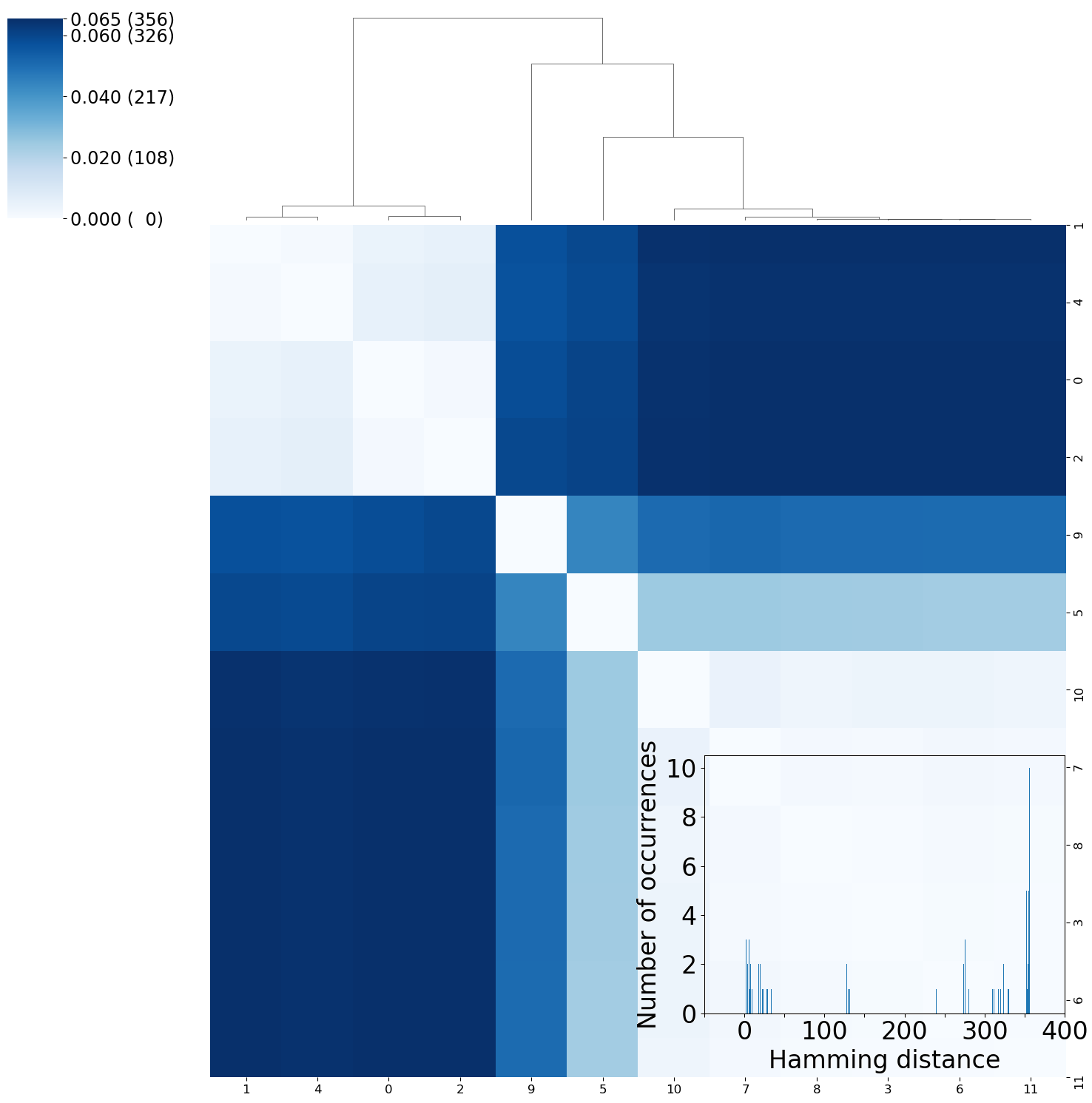}}
    \caption{With the same context as Fig. \ref{fig:diversity_2}, solutions were found in four places significantly far apart in space. Three of them were greater than a Hamming distance of 200 apart, with numerous solutions up to a distance of 350 apart.}
    \label{fig:diversity_3}
\end{figure}

Further studying the similarity of solutions vectors, we consider the distribution and cluster properties
of solutions. 
In Fig.~\ref{fig:diversity_2}-\ref{fig:diversity_3}, we selected three separate problems for which 
Mukai found the best known energy and considered the set of elite solutions 
within a relative tolerance of $10^{-6}$. 
Each subfigure in Fig.~\ref{fig:diversity_2}-\ref{fig:diversity_3} contains a matrix of distances between pairs of these elite solutions, 
a dendrogram on the top showing the hierarchical clustering of the elite solutions,
a colorbar recording Hamming distance, both normalized and (nonnormalized),
plus an inset histogram showing the distribution of distances. 
Each matrix is sorted according to the hierarchical clustering measured by its dendrogram (above the matrix), where the height of the lines in the dendrogram denotes the Hamming distance.
The histogram is computed by counting the number of pairs of distinct solutions per Hamming distance. 
For instance, for a set of solutions uniformly distributed in a small region of space, 
the histogram would approximate a normal distribution, by the law of large numbers.
Figure~\ref{fig:g001883-normal-hist} shows this distribution for a problem (g001883) 
with 700 near-equal solutions. 
A significant feature of many other problems is multiple solution regions in space 
and a relatively uniform distribution within each of these regions of space.  
Among the subplots in Fig.~\ref{fig:diversity_2}-\ref{fig:diversity_3}, the solutions differ from each other 
by Hamming distances (numbers of bit flips) ranging from one to hundreds. 
Of note, in terms of quality, they are equal to within a relative tolerance of $10^{-6}$, 
giving the user a wide range of valid solutions. 
Just as a relatively minor change in energy may yield significantly improved results, 
for applications from airline scheduling to protein folding, it is often more important 
to have a  pool of high-quality solutions, since out-of-problem factors 
such as business logic or production limitations may reduce the viability 
of a significant portion of those solutions.

\subsection{Miscellany}
The QCI qbsolv component of Mukai finds excellent energy values for many of the MQlib
instances chosen by other groups of industrial benchmarkers, and there is headroom for
improvement in its performance, notably for sparser problems.
We discovered that 9 of the selected MQlib instances appear not to meet the selection
criteria of only a single heuristic achieving the best known solution, so those
instances are not as hard as expected.\footnote{g000432, g000524, g001269, g001883, g002204, g002440,  g002586, g002898, and g003215}
Due to the limited detail in \cite{dwave2020hss}, we cannot compare our results
to those of their hybrid solver.

QCI qbsolv samples many MQlib instances very fast, in less than 25\% of the 1200-second
time limit established by MQlib.  
Still, there is room for faster time to solution by better recognition of lack of
progress.

Diversity of solutions is perhaps an underappreciated dimension of solver performance 
on which QCI qbsolv performs very well, finding as many as 150 distinct superior solutions
for one instance and finding distinct superior solutions with Hamming distances of 350
for others.  
Given QCs' exploitation of quantum effects to find widely separated solutions, having
a classical solver that emulates that diversity today is an excellent means of
being quantum-ready in an unexpected dimension.

Having explored this set of 45 MQlib instances in detail, we see that expanding our
scope to include larger MQlib problems will more accurately reflect 
the needs of industrial users needing to become quantum-ready.

As expected for an emerging technology like Mukai, there are numerous ways to
improve these early results, including examining poorly performing instances for insight into improvements,
using more classical compute resources (CPU cores or GPUs), moving to higher-level formulations
that enable better performance and ease of use, and optimizing the benefit gained
from near-term QCs via better hybrid quantum/classical methods.
We have several developments under way to deploy much greater classical resources.  
We will soon release a higher-level API that will make it easier for application
developers to specify constrained-optimization problems to Mukai and simultaneously
grant to Mukai developers more design freedom to optimize problems before 
converting them to QUBO form.
We are collaborating with an external partner on calibrating QAOA for effective
use on near-term gate-model QCs.

This work reports on solving QUBOs from a well-known public benchmark repository.
An important future step, translating this low-level benefit 
to an end-user benefit in
solving real-world problems, will be vital to establishing the reality and value
of this approach.

\section{Conclusions}
We have shown that  instances of the established QUBO formulation 
from multiple problem domains
can be solved today on classical hardware
with quantum-ready algorithms,
delivering superior results in both quality of solution and diversity of solutions. 
The approach is also affordable, delivering, in many cases, superior performance 
on affordable hardware that is readily available in public cloud services. 
As expected for an emerging technology, there are numerous ways to
improve these early results, including examining poorly performing instances,
using more classical compute resources, moving to higher-level formulations
that enable better performance and ease of use, and optimizing the benefit gained
from near-term QCs.
Still, QCI's Mukai  delivers the best known quality of results, time to
solution, and diversity of solutions in a commercially available service, 
enabling compute-dependent organizations to become quantum-ready today.

\section*{Acknowledgment}

We acknowledge the contributions that Max Dechantsreider of Performance Jones LLC made to the
performance of QCI qbsolv.






\bibliographystyle{unsrt}
\bibliography{qci_qbsolv_mqlib_perf_results}

\end{document}